\documentclass[aps, prd, nofootinbib, preprintnumbers, twocolumn, showpacs]{revtex4}
\usepackage{graphicx}
\usepackage{amsmath}
\usepackage{multirow}
\usepackage{bm}
\def\fsl#1{\setbox0=\hbox{$#1$}                 
   \dimen0=\wd0                                 
   \setbox1=\hbox{/} \dimen1=\wd1               
   \ifdim\dimen0>\dimen1                        
      \rlap{\hbox to \dimen0{\hfil/\hfil}}      
      #1                                        
   \else                                        
      \rlap{\hbox to \dimen1{\hfil$#1$\hfil}}   
      /                                         
   \fi}                                         %

\newcommand{\Tr}{\mbox{Tr}}
\newcommand{\diag}{\mbox{diag}}
\newcommand{\VEV}[1]{\langle #1 \rangle}

\begin{document}
\title{Is gluonic color-spin locked phase stable?}
\author{Michio Hashimoto}
 \email{mhashimo@uwo.ca}
  \affiliation{
   Department of Applied Mathematics, 
   University of Western Ontario, 
   London, Ontario N6A 5B7, Canada}
\date{\today}
\preprint{UWO-TH-08/03}
\pacs{12.38.-t, 11.15.Ex, 11.30.Qc}
 
\begin{abstract}
We study the gluonic color-spin locked (GCSL) phase in 
dense two-flavor quark matter.
In this phase, 
the color and spatial rotational symmetries are spontaneously
broken down to $SO(2)_{\rm diag}$ with the generator
being an appropriate linear combination of the color and 
rotational ones.
The Meissner masses of gluons and the mass of the radial mode of 
the diquark field in the GCSL phase are calculated and
it is shown that this phase is free from
the chromomagnetic and Sarma instabilities in the whole parameter region
where it exists.
The GCSL phase describes an anisotropic color and electromagnetic 
superconducting medium.
Because most of the initial symmetries in this phase are 
spontaneously broken, its dynamics is very rich.
\end{abstract}

\maketitle

It is plausible that a color superconducting phase is realized 
in the interior of compact stellar objects~\cite{review}.
Matter inside compact stars should be in $\beta$-equilibrium and 
be electrically and color neutral.
Owing to these conditions and the non-negligible strange quark mass, 
a mismatch $\delta\mu$ between the Fermi momenta of the pairing 
quarks is inevitably induced.
This is crucial for the quark-pairing dynamics~\cite{Alford:2002kj}.

Moreover, when the mismatch $\delta\mu$ increases,
the Meissner masses of gluons become imaginary
both in the gapped (2SC) and gapless (g2SC) two-flavor color 
superconducting phases~\cite{Huang:2004bg}.
A similar phenomenon has been found also in 
the three-flavor gapless color-flavor locked (gCFL) 
phase~\cite{Casalbuoni:2004tb,Alford:2005qw,Fukushima:2005cm}.

As was shown in Ref.~\cite{Gorbar:2006up},
the chromomagnetic instability at least in the 2SC/g2SC phase 
is closely connected with the appearance of tachyonic plasmons 
in the physical gluonic channels.
This supports the scenario with gluon condensates (gluonic phase) 
proposed in Refs.~\cite{Gorbar:2005rx,Gorbar:2007vx}. 

Another problem in the g2SC phase is the existence of 
the Sarma instability, 
which corresponds to the negative mass squared of 
the physical diquark excitation (the diquark Higgs mode) at zero momentum.
It was also found that the diquark Higgs mode has a negative velocity 
squared in the g2SC region~\cite{Hashimoto:2006mn}.
In Refs.~\cite{Iida:2006df,Giannakis:2006gg} 
a similar instability is discussed.

One of the urgent problems in this field is to resolve these problems.
Besides the gluonic phase~\cite{Gorbar:2005rx,Gorbar:2007vx}, 
a number of other candidates for the genuine ground state have been proposed~\cite{Alford:2000ze,Bowers:2002xr,Reddy:2004my,Huang:2005pv,Hong:2005jv,Kryjevski:2005qq,Schafer:2005ym,Casalbuoni:2005zp,Rajagopal:2006ig,Mannarelli:2007bs,Gatto:2007ja}.

In Ref.~\cite{Hashimoto:2007ut}, 
a numerical analysis for two gluonic phases, 
the minimal cylindrical gluonic phase II and 
the gluonic color-spin locked (GCSL) 
one~\cite{Gorbar:2007vx,Hashimoto:2007ut}, 
was performed\footnote{
For the earlier works, see 
Refs.~\cite{Fukushima:2006su,Kiriyama:2006ui}.
The extension to the model with nonzero temperature was studied
in Ref.~\cite{Kiriyama:2007ng}.
}.
It is shown that the gluonic phases are actually realized 
in wide regions of the parameter space and they are energetically 
more favorable than the normal, 2SC/g2SC, and 
the single plane wave Larkin-Ovchinnikov-Fulde-Ferrell 
(LOFF)~\cite{Alford:2000ze,Giannakis:2004pf,Gorbar:2005tx} phases.

A formula for the Meissner screening mass was developed 
in Ref.~\cite{Hashimoto:2007tx} and also
the Meissner masses in the minimal cylindrical gluonic 
phase~II were examined.
It was found, however, that the chromomagnetic instability is 
only partially resolved in the minimal cylindrical gluonic phase~II.
 
In this paper, 
the stability of the GCSL phase is studied.
We calculate the Meissner masses for gluons and also 
the mass of the diquark Higgs field at zero momentum.
It is shown that the GCSL phase resolves both of the chromomagnetic and
Sarma instabilities in the whole region where the phase exists.

As in Refs.~\cite{Gorbar:2005rx,Gorbar:2007vx,Hashimoto:2007ut}, 
in the analysis, the gauged Nambu-Jona-Lasinio (NJL) model with 
two light quarks will be used.
We neglect the current quark masses and
the $(\bar{\psi}\psi)^2$-interaction channel.
The Lagrangian density is then given by
\begin{eqnarray}
  {\cal L} &=& \bar{\psi}(i\fsl{D}+\bm{\mu}_0\gamma^0)\psi
  +G_\Delta \bigg[\,(\bar{\psi}^C i\varepsilon\epsilon^\alpha\gamma_5 \psi)
           (\bar{\psi} i\varepsilon\epsilon^\alpha\gamma_5 \psi^C)\,\bigg]
  \nonumber \\
&&  -\frac{1}{4}F_{\mu\nu}^{a} F^{a\,\mu\nu} ,
 \label{Lag}
\end{eqnarray}
with
\begin{equation}
  D_\mu \equiv \partial_\mu - i g A_\mu^{a} T^{a}, \quad
  F_{\mu\nu}^{a} 
  \equiv \partial_\mu A_\nu^{a} - \partial_\nu A_\mu^{a} +
  g f^{abc} A_\mu^{b} A_\nu^{c},
\end{equation}
where $\varepsilon$ and $\epsilon^{\alpha}$ are the totally 
antisymmetric tensors in the flavor and color spaces, respectively. 
We also introduced gluon fields $A_\mu^{a}$,
the QCD coupling constant $g$, 
the generators $T^{a}$ of $SU(3)$ 
and the structure constants $f^{abc}$.
The quark field $\psi$ is a flavor doublet and color triplet.
The charge-conjugate spinor is defined by
$\psi^C \equiv C \bar{\psi}^T$ with $C = i\gamma^2\gamma^0$.
We do not introduce the photon field.
On the other hand, the whole theory contains free electrons,
although we do not show them explicitly in Eq.~(\ref{Lag}).
In $\beta$-equilibrium, the chemical potential matrix $\bm{\mu}_0$ 
for up and down quarks is 
$\bm{\mu}_0 = \mu {\bf 1} - \mu_e Q_{\rm em}$, 
with ${\bf 1} \equiv {\bf 1}_c \otimes {\bf 1}_f$, and
$Q_{\rm em} \equiv {\bf 1}_c \otimes \diag(2/3,-1/3)_f$,
where
$\mu$ and $\mu_e$ are the quark and electron chemical potentials,
respectively.
(The baryon chemical potential $\mu_B$ is given by $\mu_B \equiv 3\mu$.)
The subscripts $c$ and $f$ mean that the corresponding matrices act on 
the color and flavor spaces, respectively.

\begin{figure}[t]
 \begin{center}
 \resizebox{0.47\textwidth}{!}{\includegraphics{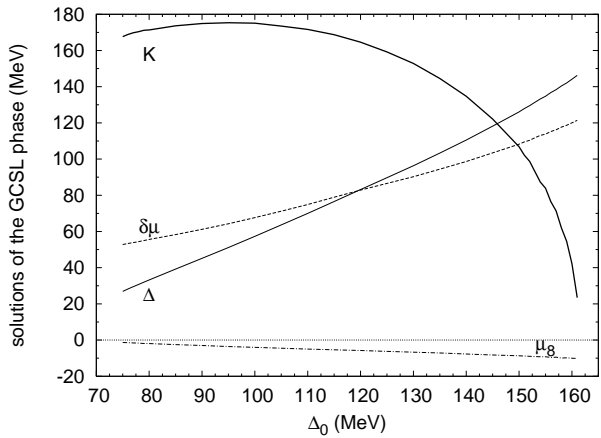}}
 \end{center}
 \caption{The dynamical solutions of the GCSL phase with $\alpha_s=1$.
          The values $\Lambda=653.3$ MeV and $\mu=400$ MeV were used.
          \label{sol_gcsl}}
\end{figure}

In the fermion one-loop approximation, 
the bare effective potential including both gluon and diquark condensates 
is given by
\begin{equation}
  V_{\rm eff}^{\rm bare} =
   \frac{\Delta^2}{4G_\Delta}
  +\frac{1}{4}F_{\mu\nu}^{a} F^{a\,\mu\nu}
  -\frac{\mu_e^4}{12\pi^2}
  -\frac{1}{2}\int\frac{d^4 P}{i(2\pi)^4}\Tr\ln S_g^{-1} ,
  \label{V_exp}
\end{equation}
where $\Delta$ and $S_g^{-1}$ denote the diquark gap and 
the fermion propagator inverse with gluon condensates
in the Nambu-Gor'kov space, respectively.
We also added the free electron contribution.
Since the bare potential has a divergence, 
a counter term is required.
The dimensional regularization is nice for gauge theories, 
but not for the NJL-type models.
We here take into account only differences of the free energies
with and without the chemical potentials.
Such effects should be physical.
This is a similar idea for the subtraction scheme
considered in Ref.~\cite{Alford:2005qw}.
Let us define the renormalized effective potential by
\begin{equation}
  V_{\rm eff}^R \equiv V_{\rm eff}^{\rm bare} - V_{\rm c.t.} ,
\end{equation}
with the counter term,
\begin{equation}
  V_{\rm c.t.} = -\frac{1}{2}\int\frac{d^4 P}{i(2\pi)^4}\Tr\ln 
  S_g^{-1} \Bigg|_{\mu=\mu_e=\mu_a=0,\Delta=0,\VEV{\vec A^a} \ne 0} \, . 
  \label{ct}
\end{equation}
In this prescription, even if we use the regularization scheme 
with the sharp three-momentum cutoff $\Lambda$ for the loop integral, 
we can remove artificial mass terms of gluons 
like $\Lambda^2 \vec A_a^2$.

\begin{figure}[t]
 \begin{center}
 \resizebox{0.47\textwidth}{!}{\includegraphics{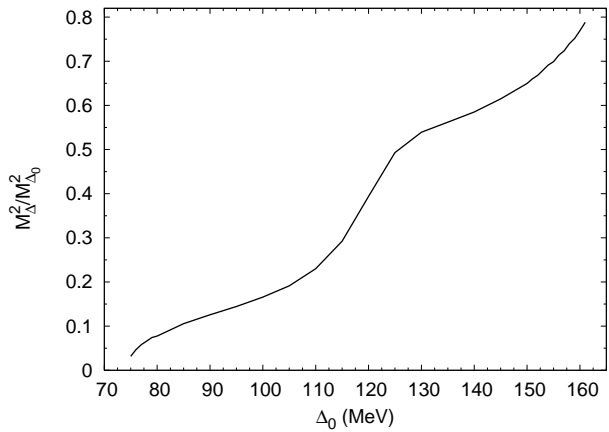}}
 \end{center}
 \caption{The mass of the diquark Higgs field in the unit of
          $M_{\Delta_0}^2(\equiv 4\mu^2/\pi^2)$.
          The values $\Lambda=653.3$ MeV, $\mu=400$ MeV and
          $\alpha_s=1$ were used.
          \label{md}}
\end{figure}

In this paper, we study the GCSL 
phase~\cite{Gorbar:2007vx,Hashimoto:2007ut} with an ansatz for 
the gluon condensates,
\begin{equation}
  \mu_8 \equiv \frac{\sqrt{3}}{2} g\VEV{A_0^{8}}, \quad
  K \equiv g \VEV{A_x^{4}} = g \VEV{A_z^{6}} \, .
\end{equation}
The dynamics of the GCSL phase was
analyzed in Ref.~\cite{Hashimoto:2007ut}.
In order to find the dynamical solutions, 
we searched for minima of the (neutral) effective potential 
imposed the neutrality conditions.
We also converted the four-diquark coupling constant $G_\Delta$
to the 2SC gap parameter $\Delta_0$ defined at $\delta\mu=0$
and varied the values of $\Delta_0$ from the weak coupling regime
($\Delta_0 \sim 60$ MeV) to the strong coupling one
($\Delta_0 \sim 160$ MeV).
The dynamical solutions of the GCSL phase are shown in Fig.\ref{sol_gcsl}.
In the analysis, we took realistic values $\mu=400$MeV and
$\Lambda=653.3$MeV.
For the gluonic phase, it is required to specify the value of 
$\alpha_s [\equiv g^2/(4\pi)]$, because of the existence of 
the tree gluon potential term.
In Ref.~\cite{Rischke:2000cn}, the confinement scale is estimated 
to be not so large.
In Fig.\ref{sol_gcsl}, we took $\alpha_s=1$ as a typical value.
We then find that the GCSL phase exists in the region
\begin{equation}
    \mbox{76 MeV} < \Delta_0 < \mbox{161 MeV}, 
\end{equation}
for $\alpha_s=1$.
While the values of $K$ tend to be small but nonzero 
near the endpoint of the GCSL phase, $\Delta_0 \sim 161$ MeV,
the gap $\Delta$ and the gluon condensate $K$ are still big
even in the vicinity of the starting point, 
$\Delta_0 \sim 76$ MeV. (See Fig.\ref{sol_gcsl}.) 
These suggest that the first order phase transition
occurs both at the starting and end points of the GCSL phase.

Note that the values of $\Delta_0$ 
at the starting point of the GCSL phase are rather sensitive 
to the choice of $\alpha_s$, because the tree gluon potential term 
is non-negligible, i.e., as the values of $\alpha_s$ are bigger,
the windows are wider~\cite{Hashimoto:2007ut}.

For reasonable parameter regions, say,
$\mbox{300 MeV} < \mu < \mbox{500 MeV}$ and
$653.3\mbox{ MeV} < \Lambda < \mbox{1 GeV}$,
the behaviors of the solutions are qualitatively unchanged.
When we vary the values of $\mu$ from $300$ MeV 
to $400$ MeV and from $400$ MeV to $500$ MeV,
the windows where the GCSL phase is energetically stabler than 
the minimal cylindrical gluonic phase II are about 15\% wider, 
respectively.
On the other hand, when the value of $\Lambda$ is taken 
to $\Lambda=1$ GeV from $\Lambda=653.3$ MeV, 
the window is about 30\% wider.

We now analyze the Meissner screening masses and 
the mass of the diquark Higgs field in the GCSL phase by using
the formulae derived in Ref.~\cite{Hashimoto:2007tx},
\begin{widetext}
\begin{eqnarray}
  \frac{\partial^2 V_{\rm eff}^R}{\partial \Delta^2} &=&
   \frac{1}{2G_\Delta}
  -\frac{1}{2}\sum_{\tau=\pm}\sum_{E_i^\tau \ne E_j^\tau}\int\frac{d^3 p}{(2\pi)^3}
    \frac{\theta(E_i^\tau)-\theta(E_j^\tau)}{E_i^\tau-E_j^\tau}
      (U^\dagger \tilde{\Gamma}_{\Delta} U)_{ij}
      (U^\dagger \tilde{\Gamma}_{\Delta} U)_{ji} \nonumber \\
&&-\frac{1}{2}\sum_{\tau=\pm}\sum_{E_i^\tau=E_j^\tau}\int\frac{d^3 p}{(2\pi)^3}
      \delta(E_i^\tau)
      (U^\dagger \tilde{\Gamma}_{\Delta} U)_{ij}
      (U^\dagger \tilde{\Gamma}_{\Delta} U)_{ji} ,
  \label{d2V3-1}
\end{eqnarray}
and
\begin{eqnarray}
  \frac{\partial^2 V_{\rm eff}^R}{\partial A_\mu^a \partial A_\nu^b} &=&
   \Pi^{\mu\nu}_{\rm tree}
 -\frac{g^2}{2}\sum_{\tau=\pm}\sum_{E_i^\tau \ne E_j^\tau}\int\frac{d^3 p}{(2\pi)^3}
    \frac{\theta(E_i^\tau)-\theta(E_j^\tau)}{E_i^\tau-E_j^\tau}
      (U^\dagger \tilde{\Gamma}^{\mu a} U)_{ij}
      (U^\dagger \tilde{\Gamma}^{\nu b} U)_{ji} \nonumber \\
&&-\frac{g^2}{2}\sum_{\tau=\pm}\sum_{E_i^\tau=E_j^\tau}\int\frac{d^3 p}{(2\pi)^3}
      \delta(E_i^\tau)
      (U^\dagger \tilde{\Gamma}^{\mu a} U)_{ij}
      (U^\dagger \tilde{\Gamma}^{\nu b} U)_{ji} 
  \; -\mbox{(counter term)} ,
  \label{d2V3-3}
\end{eqnarray}
\end{widetext}
where $E_{1,2,\cdots,n}^\tau$ denote the energy eigenvalues of
$S_g^{-1}$.
We also defined the tree contribution 
\begin{eqnarray}
  \Pi^{\mu\nu}_{\rm tree} &\equiv& \phantom{+}
    g^2 f^{a_1 a b}f^{a_1 a_2 a_3} A^{a_2\,\mu} A^{a_3\,\nu}
   \nonumber \\
&&+ g^2 f^{a_1 a a_2}f^{a_1 b a_3} g^{\mu\nu} A_\lambda^{a_2} A^{a_3\,\lambda}
   \nonumber \\
&&+ g^2 f^{a_1 a a_2}f^{a_1 a_3 b} A^{a_3\,\mu} A^{a_2\,\nu},
\end{eqnarray}
and the transformed vertices
\begin{equation}
  \tilde{\Gamma}_\Delta = \epsilon^b
  \left(\begin{array}{cc}
   0 & 1 \\ -1 & 0
  \end{array}\right) ,
\end{equation}
and
\begin{equation}
  \tilde{\Gamma}^{\mu a} = 
  \left(\begin{array}{cc}
   \gamma^0 \gamma^\mu T^a & 0 \\ 0 & -\gamma^\mu \gamma^0 (T^a)^T
  \end{array}\right) ,  
\end{equation}
in the Nambu-Gor'kov space.
For the Dirac's $\delta$-function, a lot of expressions are known.
In our calculation, we use 
$\delta(x)={\displaystyle \lim_{n\to\infty}}\sqrt{n/\pi}\exp(-nx^2)$.
We also crosscheck our calculation by using the numerical derivative
of the effective potential.

Let us introduce the notations
\begin{equation}
  M^2_\Delta \equiv 
  \frac{\partial^2 V_{\rm eff}^R}{\partial \Delta^2}, \qquad
  (M^2)_{a_i b_j} \equiv 
  \frac{\partial^2 V_{\rm eff}^R}{\partial A_i^a \partial A_j^b} ,  
\end{equation}
and 
\begin{equation}
    (M^2)_{a_i\mbox{-}b_j,n}, \qquad
    (M^2)_{a_i\mbox{-}b_j\mbox{-}c_k,n} \; \; ,
\end{equation}
for the eigenvalues of the Meissner mass 
($2\times 2$ and $3 \times 3$) matrices, 
where $(M^2)_{a_i\mbox{-}b_j,1} \leq (M^2)_{a_i\mbox{-}b_j,2}$
and so on.
Notice that there appear mixing terms among the space components of gluons
owing to the symmetry breaking structure of 
the GCSL phase~\cite{Gorbar:2007vx}
\begin{equation}
 SU(3)_c \times U(1)_{\rm em} \times SO(3)_{\rm rot} 
 \stackrel{\Delta,K}{\longrightarrow}
  SO(2)_{\rm diag},
\end{equation}
where the unbroken $SO(2)_{\rm diag}$ symmetry consists of 
an appropriate linear combination of the initial color and 
rotational symmetries.
At least the mixing terms which exist in the tree contribution
should be taken into account.
To reduce our labor, 
we might ignore the other types of the mixing terms.

\begin{figure}[t]
 \begin{center}
 \resizebox{0.47\textwidth}{!}{\includegraphics{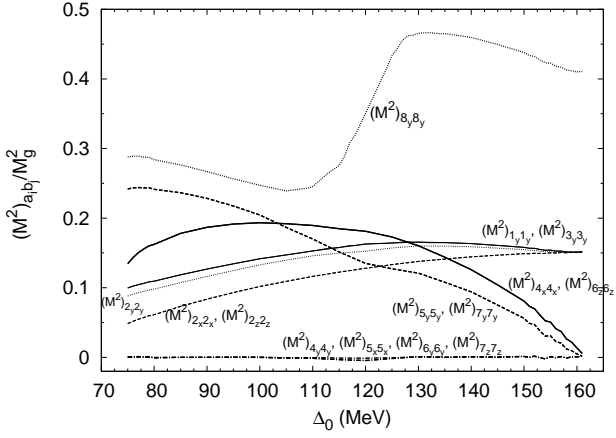}}
 \end{center}
 \caption{The Meissner masses for the GCSL phase without the mixing terms
          in the unit of $M^2_g [\equiv 4\alpha_s \mu^2/(3\pi)]$.
          The values $\Lambda=653.3$ MeV, $\mu=400$ MeV and
          $\alpha_s=1$ were used.
          \label{m_fig1}}
\end{figure}

\begin{figure}[t]
 \begin{center}
 \resizebox{0.47\textwidth}{!}{\includegraphics{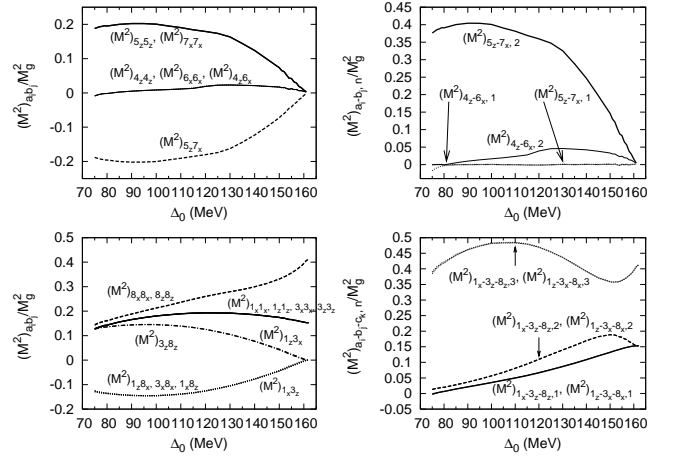}}
 \end{center}
 \caption{The Meissner masses for the GCSL phase with the mixing terms
          in the unit of $M^2_g [\equiv 4\alpha_s \mu^2/(3\pi)]$.
          The values $\Lambda=653.3$ MeV, $\mu=400$ MeV and
          $\alpha_s=1$ were used.
          \label{m_fig2}}
\end{figure}

We depict the numerical results in Figs.~\ref{md}--\ref{m_fig2}
in the units of $M_{\Delta_0}^2 \equiv 4\mu^2/\pi^2$ and 
$M^2_g \equiv 4\alpha_s \mu^2/(3\pi)$.
In the whole region where the GCSL phase exists, 
it turns out that the diquark Higgs mass and
all of the eigenvalues of the Meissner masses are positive.
We also find that the relations 
$(M^2)_{1_x1_x} \simeq (M^2)_{1_z1_z} \simeq (M^2)_{3_x3_x} \simeq (M^2)_{3_z3_z}$,
$(M^2)_{1_y1_y} \simeq (M^2)_{3_y3_y}$,
$(M^2)_{1_x3_z} \simeq -(M^2)_{1_z3_x}$,
$(M^2)_{1_z8_x} \simeq (M^2)_{1_x8_z} \simeq (M^2)_{3_x8_x} \simeq -(M^2)_{3_z8_z}$,
$(M^2)_{2_x2_x} \simeq (M^2)_{2_z2_z}$, 
$(M^2)_{4_x4_x} \simeq (M^2)_{6_z6_z}$, 
$(M^2)_{4_z4_z} \simeq (M^2)_{6_x6_x} \simeq (M^2)_{4_z6_x}$, 
$(M^2)_{5_y5_y} \simeq (M^2)_{7_y7_y}$,
$(M^2)_{5_z5_z} \simeq (M^2)_{7_x7_x} \simeq -(M^2)_{5_z7_x}$ and 
$(M^2)_{8_x8_x} \simeq (M^2)_{8_z8_z}$ numerically hold. 
The behavior of $(M^2)_{8_y8_y}$ in Fig.\ref{m_fig1} looks peculiar.
Although the numerical calculation itself is robust, because
the behavior is unchanged by varying the values of $\alpha_s$, $\mu$
and $\Lambda$, the reason is unclear at present.
It is noticeable that there are six vanishing masses
within the numerical precision, 
$(M^2)_{4_y 4_y}, (M^2)_{5_x 5_x}, (M^2)_{6_y 6_y}, (M^2)_{7_z 7_z}$
and $(M^2)_{4_z\mbox{-}6_x,1}, (M^2)_{5_z\mbox{-}7_x,1}$.
Three among the six should correspond to the gauge fixing fields.
One might guess the other three should be connected with 
the Nambu-Goldstone (NG) bosons owing to the global symmetry
breaking $U(1) \times SO(3) \to SO(2)$.
(Note that we did not introduce the photon field.)
This is unclear, however, because the kinetic terms are quite 
important for the abnormal number of the NG bosons~\cite{Miransky:2001tw}.
This question will be answered elsewhere.

Since the tree contributions are proportional to $K^2$,
the $\alpha_s$-dependence of the Meissner masses are 
numerically non-negligible.
Nevertheless, 
the results are qualitatively unchanged for $0.75 < \alpha_s < 1.25$.
We have also checked positivity of the Meissner masses and 
the diquark Higgs mass in the reasonable parameter regions, 
$\mbox{300 MeV} < \mu < \mbox{500 MeV}$ and
$653.3\mbox{ MeV} < \Lambda < \mbox{1 GeV}$.

In summary, 
we analyzed the Meissner screening masses and 
the mass of the diquark Higgs field in the GCSL phase.
We showed that unlike the minimal cylindrical gluonic phase II
the GCSL phase resolves both of the chromomagnetic and Sarma
instabilities in the whole parameter region where the phase exists.
In this sense, the GCSL phase describes a stable vacuum
against the fluctuations of the diquark and gluonic channels.
These results are qualitatively unchanged 
for realistic values of $\alpha_s$, say, $0.75 < \alpha_s < 1.25$,
and for the reasonable parameter regions, 
$\mbox{300 MeV} < \mu < \mbox{500 MeV}$ and
$653.3\mbox{ MeV} < \Lambda < \mbox{1 GeV}$.

Why can the GCSL phase resolve the chromomagnetic instability?
In the minimal cylindrical gluonic phase II only with 
$B \equiv \VEV{A_z^6} \ne 0$, the squared Meissner mass of 
the transverse mode of the 4th gluon becomes negative 
at a certain value of $\Delta_0$~\cite{Hashimoto:2007tx}.
Therefore, the GCSL phase with $\VEV{A_z^6} \ne 0$ {\it and} 
$\VEV{A_x^4} \ne 0$ can be energetically stabler than 
the minimal cylindrical gluonic phase II and 
resolve this instability.
Actually, we have found that the GCSL phase is energetically 
more favorable over the minimal cylindrical gluonic phase II 
and the single plane wave LOFF phase in a wide parameter 
region~\cite{Hashimoto:2007ut}.
After the rearrangement of the vacuum, 
the curvatures of the effective potential, i.e., the Meissner masses 
for $A_x^4$ and $A_z^6$, are likely to be positive.
Moreover, unlike in the minimal cylindrical gluonic phase II,
all of the gluonic fluctuations except for $A_z^4$ and $A_x^6$ 
have big tree terms proportional to $K^2$.
Thus the squared Meissner masses tend to be positive.

We also comment on the discrepancy between 
the starting points of the GCSL phase and
the chromomagnetic instability of $A_{x,y}^4$ in the minimal cylindrical 
gluonic phase II~\cite{Hashimoto:2007ut,Hashimoto:2007tx}.
The structures of the tree gluon potential between the two phases
are different.
In addition, the values of the dynamical solutions $B$ and $K$ are 
the same order, but numerically disagree~\cite{Hashimoto:2007ut}.
It is thus possible that the discrepancy numerically occurs.

The GCSL phase has other noticeable features;
(1) Both of the chromoelectric and chromomagnetic 
fields are spontaneously generated.
(2) The GCSL phase describes an anisotropic medium.
(3) The medium is a color and electromagnetic superconductor.

In the confinement picture, the symmetry breaking structure of
the GCSL phase and the cylindrical gluonic phase I is 
the same~\cite{Gorbar:2007vx}.
How about the relation between them?
In addition, the question whether or not the abnormal number of 
the NG bosons occurs is still open.
Last but not least, is the GCSL phase the global vacuum
in dense two-flavor quark matter?
We hope to return to these problems elsewhere.

\acknowledgments
The author thanks V.~A.~Miransky for fruitful discussions.
This work was made possible by the facilities of the Shared Hierarchical
Academic Research Computing Network (SHARCNET:www.sharcnet.ca).
The research was supported by the Natural Sciences and Engineering
Research Council of Canada.

\end{document}